\documentclass[twocolumn,pra,showpacs,superscriptaddress,amssymb,amsmath,amsmath,floatfix]{revtex4-1}

\usepackage{graphicx}
\usepackage{epstopdf}
\usepackage{bm}
\usepackage[colorlinks=true,linkcolor=blue,urlcolor=blue,citecolor=blue]{hyperref}
\usepackage{comment}
\usepackage{float}
\usepackage{cancel}
\usepackage{times}
\usepackage{amsmath, amsthm, amssymb, amsfonts}
\usepackage{xcolor}

\newcommand{\affFUW}{Faculty of Physics, University of Warsaw, Pasteura 5, 02-093 Warsaw, Poland}

\begin{document}
\author{Krzysztof Myśliwy and Krzysztof Jachymski}
\affiliation{\affFUW}
\date{\today}
\title{Simulating critical anion chemistry with indirect excitons}
 \begin{abstract}
 	We study the effect of geometric constraints on the formation of indirect excitonic complexes with excess charge by considering the problem of two identical electrons moving in half-space subject to mutual Coulomb repulsion and the Coulomb attraction to a heavy hole of charge $Z$ residing on a flat impenetrable surface. We find that $Z=1$ is the critical charge necessary for binding, as compared to $Z_c\approx0.911$ in the unconstrained system. This suggests that interlayer electron--hole systems may serve as a flexible and experimentally accessible platform to study electronic behavior close to the critical nuclear charge necessary for binding.
 \end{abstract}
\maketitle
\section{Introduction}

Quantum mechanics is able not only to explain the stability of atoms and molecules, but also to predict their structure with essentially arbitrary accuracy, as long as computational resources are sufficient. These predictions agree well with experiment, and the applicability of computational and theoretical methods covers virtually all conceivable materials, no matter how complex, and it is there where most of the theoretical impetus goes. Given this tremendous success, it is easy to overlook a handful of very basic problems in quantum chemistry that still lack sufficient theoretical understanding. One of them is the following: how much excess electrons can a nucleus of charge $Z$ bind, and why does it appear to be at most one or two, regardless of $Z$? This problem is known among mathematicians as the \emph{maximal ionization problem} \cite{Nam2022} and while there has been some progress in giving an answer, often accompanied by profound developments in mathematics  \cite{Nam2012, Fefferman1990, Solovej2003, Lieb1984} and including key physics insights \cite{LSST1984}, the question remains unresolved. 

The most important reason for the inherent difficulty of such an apparently simple question is the role played by electronic correlation in the stability of such systems. Since electron correlation is enormously important in many aspects of chemistry and materials science, this provides further motivation to study the maximal ionization problem, with the hope that new insight on electron correlation might emerge from the solution. From the theoretical perspective, the question has been turned around a bit: instead of fixing the charge of the nucleus $Z$, one fixes the number of electrons $N$ and varies $Z$, treating it as a continuous parameter, and looks for the minimal value of $Z$ that leads to binding. For example, for the helium isoelectronic sequence $N=2$, the critical value of $Z$ necessary for binding has been determined to a high degree of accuracy, and assumes the value $Z_c=0.91102822407725573(4)$ in the infinite nucleus mass approximation \cite{Estienne2014, Baker1990} (see \cite{Kais2000} for the full phase diagram involving the variation of masses, and \cite{Pachucki2012} for the case $N=3$). In particular, there exists one stable species with excess negative charge in this sequence, namely the hydrogen anion H$^-$, whose existence was first predicted by Bethe in \cite{Bethe1929}  with later works on its binding energy by Chandrasekhar \cite{Chandrasekhar1944} who was motivated by the role played by this anion in stellar spectroscopy \cite{Rau1996}. It can be shown that for $Z=1$, two is the maximal number of electrons that can be bound \cite{Lieb1984}.

In 1958, Lampert \cite{Lampert1958} was first to discuss the formation of bound complexes made up of charged particles in solids, thus extending the concept of the exciton to many- particle structures analogous to ordinary chemical species. Among these complexes, the simplest one is the trion, a three-body bound state comprising of two equally charged particles bound to a single oppositely charged one. This system is an analog of the negative hydrogen ion. A particularly interesting case is the one where the charges are constrained to different layers of a composite material. Such complexes, typically called indirect or interlayer excitons, are currently a very active field of research~\cite{Wang2018}. Since the geometry imposes a permanent dipole moment on such a complex, the indirect excitons thus formed are expected to interact strongly and if the system is sufficiently well controllable it can be viewed as a quantum simulator of strongly correlated physics. In recent years, excitons trapped in layered heterostructures have indeed become useful in this context~\cite{Rafal2022,Taylor2022,Lagoin2022,Gu2022}. In particular, electron-exciton interactions have been studied and shown to exhibit Feshbach resonance physics allowing for tuning the scattering length~\cite{Fey2020,Efimkin2021,Schwartz2021}. The geometry of the system can also be varied from quasi-1D to quasi-2D with variable number of layers~\cite{Schinner2013,Slobodkin2020,Thureja2022}. Studying the properties of few-body systems involving electrons, holes and their bound complexes in a structured environment is then not only of fundamental interest but can be relevant for understanding experimental data~\cite{Ganchev2015,Varga2020}.

In this work, we propose to use the excitonic system to simulate the critical binding mechanism in atoms at the edge of stability, i.e. close to $Z_c$, which is hard to reproduce directly, thus turning around the popular paradigm of atomic quantum simulators for condensed matter phenomena. Given that under normal circumstances $Z_c$ is quite far from the closest integer, so far such investigations have necessarily been limited to theory. We are going to put forward the idea that this stability edge could possibly be examined experimentally with indirect excitonic complexes employed to simulate the chemical systems in question. In fact, we are going to show that in the idealized model of an indirect exciton formed in two layers of disparate thickness, where the thin layer is rich in heavy holes, the critical charge of the hole needed to bind two electrons is equal to unity, as the geometry of the sample imposes additional electronic correlation, raising the value of $Z$ necessary for binding. Therefore, weakly bound trions formed in systems with similar geometries bear certain resemblance to anions at the binding stability edge. In this way, experimental investigation of such complexes may provide some information on critical anion chemistry, and possibly provide some fresh insight on the maximal ionization problem. 

This work is structured as follows. In Section~\ref{sec:setting} we discuss the structure of the setup we have in mind and the properties of a single electron interacting with a hole leading to indirect exciton state. Then in Section~\ref{sec:main} we proceed to the geometrically restricted trions employing the Chandrasekhar ansatz. The implications of the results are discussed in Sec.~\ref{sec:discuss}.

\section{Setting}
\label{sec:setting}
\begin{figure}
\includegraphics[scale=0.4]{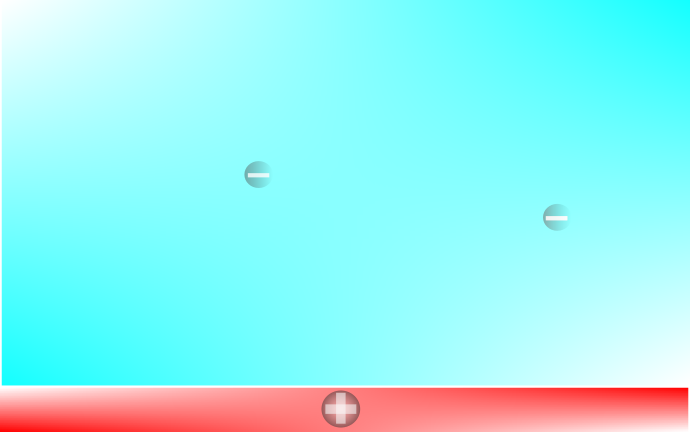}
\caption{Schematic depiction of the arrangement considered.}\label{fig1}
\end{figure}

The setting that we want to study is depicted in Fig. \ref{fig1}.
%Two materials, one rich in electrons and the other rich in holes are arranged into layers of thickness $d_1$, $d_2$ respectively, and kept at a distance $d_0$ apart.
A quasi--2d material hosting a hole is placed in the vicinity of another material of finite thickness, which hosts two identical electrons. We assume that the effective mass of the hole is much larger than the effective mass of the electrons and work in the infinite--mass approximation. Moreover, we neglect the finite size effects of the electron--rich material and treat it as a half--space. The charges interact directly via electrostatic forces, but in general their effective interaction is not of the pure Coulomb type and is modified due to dielectric effects imposed by the boundaries, as is usually described by the Rytova-Keldysh potential~\cite{Tuan2018}. The formation and stability of bound complexes formed with interactions ranging across boundaries has been already studied in some cases, in particular the trion formation in the limit where both the electon--rich and the hole--rich materials are essentially $2d$  and kept at a distance $d$ apart \cite{Witham2018}. In that scenario the trion can be formed provided that $d_0$ is small enough, and in particular it exists when the particles are hosted by a single quasi $2d$ system, although the Coulomb interaction is then modified considerably. Here, we are going to study the limit of layers of incommensurate thickness as sketched. We also neglect the dielectric effects and use the pure Coulomb potential. Apart from the simplicity of this geometry, the fact that the ratio of the thicknesses of the “electronic” to the “hole” layer is very large has the advantage that in this arrangement, the behaviour of the electrons is similar to the one in atomic systems with $P$ orbital symmetry, as we shall show below. The corrections stemming from dielectric boundary effects will be shortly discussed in Sec.~\ref{sec:discuss}.

In appropriate atomic-like units where $\hbar=m_e=1$, with $m_e$ the effective mass of the electrons, the problem is approximately described by the Hamiltonian 
\begin{equation}
\mathbb{H}_2=
\begin{cases}
-\frac{1}{2}\Delta_{\vec{r}_1}-\frac{1}{2}\Delta_{\vec{r}_2}-\frac{Z}{r_1}-\frac{Z}{r_2}+\frac{1}{|\vec{r}_1-\vec{r}_2|} & z_1\&  z_2\geq 0 \\
\infty, &  \text{otherwise}
\end{cases}
\end{equation}
where $Z>0$ is the effective charge of the stationary hole, $\vec{r}_1,\vec{r}_2$ denote the positions of the electrons, with $r_i\equiv|\vec{r}_i|$, and $\Delta_x$ denotes the Laplace operator. The interlayer boundary has been put on the plane $z=0$. In what follows, we will be interested in the ground state energy of $\mathbb{H}_2$ as a function of $Z$, especially in comparison with the ground state energy of the corresponding one--electron problem, i.e., 
\begin{equation}
\mathbb{H}_1=
\begin{cases}
-\frac{1}{2}\Delta_{\vec{r}_1}-\frac{Z}{r_1} & z_1\geq 0 \\
\infty, &  z_1<0
\end{cases}
\end{equation}
which models indirect interlayer excitons in the considered geometry. In order to understand the binding of the indirect trion state, we first need to discuss the indirect exciton problem described by $\mathbb{H}_1$, which we shall briefly do next.

\subsection{One--electron problem}
One can find the eigenstates of $\mathbb{H}_1$ by simply solving the usual Schrödinger equation for the hydrogen atom with Dirichlet boundary conditions on the plane $z=0$. Thus, the eigenstates of this problem are the eigenfunctions of the hydrogen atom which vanish at the plane $z=0$. If we adopt the standard spherical coordinates, then we clearly need to look for spherical harmonics that vanish for $\theta=\frac{\pi}{2}$, which is equivalent to finding pairs $(l,m)$ for which the associated Legendre polynomials $P^m_{l}(x)$ have $0$ as a root. We start with the identity valid for $m\geq 0$: 
\begin{equation}\label{leg}
P^m_l(x)=(-1)^m 2^l (1-x^2)^{\frac{m}{2}}\sum_{k=m}^l \frac{k!}{(k-m)!}x^{k-m}\binom{l}{k}\binom{\frac{l+k-1}{2}}{l}
\end{equation}
with the generalized binomial coefficient defined for any $t\in \mathbb{R}$ as
\begin{equation}
\binom{t}{k}=\frac{t(t-1)\cdots(t-k+1)}{k!}.
\end{equation} 
%The above representation can be easily verified by checking that it satisfies the defining equation for the associated Legendre polynomials 
%\begin{equation}
%P^m_l(x)....
%\end{equation}
From \eqref{leg}, it follows that 
\begin{equation}
P^m_l(0)=(-1)^{\frac{3m}{2}}\frac{2^{l-m+1}}{(l-m)!}\prod_{i=0}^{m-1}(l+m-2i-1)
\end{equation}
and, consequently, that 
\begin{equation}
P^m_l(0)=0 \iff l+m=2i+1, i=0,\cdots, m-1. \quad (m\geq 0). 
\end{equation}
Thus, the eigenstates of $\mathbb{H}_1$ are given by
\begin{equation}
\psi_{nlm}=
\begin{cases}
\sqrt{2}|nlm\rangle & z\geq 0\\
0 & z<0
\end{cases}
\end{equation}
 where $|nlm\rangle$ are the corresponding eigenstates of the hydrogen--like atom with charge $Z$, and the quantum numbers satisfy
 \begin{equation}\label{nms}
 \begin{split}
& n\geq 2, \\
& 1\leq l\leq n-1,\\
& |m|\leq l ,\\
& |m|+l \in 2\mathbb{Z}+1.
\end{split}
 \end{equation}
In particular, the ground state, of energy $-\frac{Z^2}{8}$, reads 
%\begin{equation}\label{gs}
%|2p_0^+; Z\rangle=\frac{Z^{5/2}}{8\sqrt{\pi}}(z+|z|) e^{-\frac{Zr}{2}}
%\end{equation}
\begin{equation}\label{gs}
\psi_{2p_0^+}(r; Z)=\frac{Z^{5/2}}{8\sqrt{\pi}}(z+|z|) e^{-\frac{Zr}{2}}
\end{equation}
The eigenenergies are still $-\frac{Z}{2n^2}$ as for the hydrogen--like atom but their enumeration starts at $n=2$, and their degeneracy is reduced from $n^2$ to $\frac{n(n-1)}{2}$, as quickly follows from~\eqref{nms}, in particular the angular momentum spaces contain only $l$ admissible states with a definite projection onto the quantization axis instead of the usual $2l+1$. For instance, the excited state is triply degenerate, with the "half--orbitals"  $3p_0, 3d_1, 3d_{-1}$. The classification given here is important for further analysis, but also can be generally relevant for understanding the excited states of interlayer excitons composed of mobile electrons and holes.

\section{Two--electron problem -- Chandrasekhar Ansatz}
\label{sec:main}
We now turn to the two--electron problem described by $\mathbb{H}_2$. Based on the results of the preceding section, the electron configuration of the supposed bound state should be of the type $2p_0^2$ and correspond to a singlet, since there is only one one--electron state of the appropriate symmetry available in the lowest energy sector. 

First, let us observe that a bound state of our problem in the singlet sector is excluded in the case $Z<1$: in fact, such a state could be employed to construct an excited bound state of the hydrogenic anion in the singlet configuration which, as is well known, does not exist~\cite{Hill1977}.
Thus, to show that $Z=1$ is critical it suffices to find a trial state in the singlet configuration which leads to binding for $Z>1$. We are going to build it upon the trial state introduced by Chandrasekhar~\cite{Chandrasekhar1944} which has a simple structure with a clear physical interpretation, which we generalize to our setting in a straightforward way%\begin{equation}
%\Psi(\vec{r_1},\vec{r_2};\alpha,\beta)=
%\begin{cases}
%N_{\alpha,\beta}r_1r_2 \cos(\theta_1)\cos(\theta_2)\left(e^{-\alpha r_1}e^{-\beta r_2}+e^{-\alpha r_2}e^{-\beta r_1}\right)&  0\leq \theta_i %\leq \frac{\pi}{2};\\
%0 &  \frac{\pi}{2}\leq \theta_i \leq\pi
%\end{cases}
%\end{equation}
%\begin{widetext}
%\begin{equation}\label{ansatz}
%\Psi(\vec{r_1},\vec{r_2};\alpha,\beta)=\frac{\sqrt{2}}{\pi}\frac{\alpha^{\frac{5}{2}}\beta^{\frac{5}{2}}}{\sqrt{1+\left(\frac{2\sqrt{\alpha\beta}}{\alpha+\beta}\right)^{10}}}\
%\left(z_1+|z_1|\right)\left(z_2+|z_2|\right)\left(e^{-\alpha r_1}e^{-\beta r_2}+e^{-\alpha r_2}e^{-\beta r_1}\right)
%\end{equation}
%\end{widetext}
\begin{equation}\label{ansatz}
\begin{split}
\Psi_{\alpha,\beta}(\vec{r_1},\vec{r_2})=&\frac{2\sqrt{2}}{\sqrt{1+\left(\frac{2\sqrt{\alpha\beta}}{\alpha+\beta}\right)^{10}}} \left(\psi_{2p_0^+}(r_1; 2\alpha)\psi_{2p_0^+}(r_2; 2\beta)\right. \\&+\left. \psi_{2p_0^+}(r_2; 2\alpha)\psi_{2p_0^+}(r_1; 2\beta)\right)
\end{split}
\end{equation}
with $\psi_{2p_0^+}$ defined in \eqref{gs}. 
The function is normalized,  symmetric in the electronic coordinates (recall that we assume that the electrons are in a singlet configuration), satisfies the boundary condition $\Psi(z_1=0,z_2=0)=0$ and depends on two non--linear variational parameters $\alpha,\beta$ whose inverses are interpreted as the effective Bohr radii of the electrons.The Ansatz is able to capture the effects of electron repulsion by adjusting the values of $\alpha, \beta$ while preserving the $2P_0$ symmetry of the ground state imposed by the boundary, and is thus said to capture the \emph{radial correlation}~\cite{Rau1996}.  Moreover, it is also able to describe the unbound states with one of the electrons escaping to infinity as an essentially free particle, when one of the parameters, say $\alpha$, tends to zero while the other assumes the value for the indirect exciton $\beta=0.5Z$.
% More precisely, as $\alpha\ll 1$, 
%\begin{equation}
%\Psi_{\alpha,\beta}(\vec{r_1},\vec{r_2})\sim \frac{1}{\sqrt{2}}\left(|2p_0^+\rangle\otimes |1,0,0\rangle +|1,0,0\rangle\otimes |2p_0^+\rangle|\right)
%\end{equation}
%with $(|2p_0^+\rangle$, defined as in \eqref{gs},
%is the ground state of the system composed of one electron bound to a heavy hole of charge $Z$ residing on a flat surface impenetrable to the electron, and $|1,0,0\rangle \sim \alpha^{5/2}z$ is the asymptotic behavior of the ground state of a free particle in a very large (compared to the scale set by the Bohr radius) cylindrical box of surface area $\sim \alpha^{-2}$ and height $\alpha^{-1}$ with Dirichlet boundary conditions at the bottom base. 
Accordingly, it can be expected that the Ansatz correctly describes the electrons close to the stability edge where one of the effective Bohr radii is very large. Moreover, thanks to its simplicity, it allows for a relatively transparent and easy to interpret description of the electronic behavior at the stability edge.  We are going to shortly discuss its limitations by considering a simple extensions of \eqref{ansatz} with correlated angles in Sec. ~\ref{sec:discuss}.

\subsection{Expected values}
The relevant one--electron integrals of $\Psi(\vec{r_1},\vec{r_2};\alpha,\beta)$ are straightforward to evaluate.
%In particular, the norm is 
%\begin{equation}
%N_{\alpha,\beta}^{-1}=\frac{1}{H_{\alpha,\beta}^{10}}+\frac{1}{A_{\alpha, \beta}^{10}}
%\end{equation}
%where $H_{\alpha,\beta}=\sqrt{\alpha\beta}$, $A_{\alpha,\beta}=\frac{\alpha+\beta}{2}$ are the geometric and arithmetic means, respectively, of $%\alpha,\beta$. 
The kinetic energy equals 
\begin{equation}
T_{\alpha,\beta}=\frac{1}{2}\left(\alpha^2+\beta^2\right)+\alpha\beta \frac{\left(\frac{2\sqrt{\alpha\beta}}{\alpha+\beta}\right)^{10}}{1+\left(\frac{2\sqrt{\alpha\beta}}{\alpha+\beta}\right)^{10}};
\end{equation}
radial correlation is here manifest in the second term. 
The expected value of the electron-hole interactions is simply
\begin{equation}
V_{\alpha,\beta,Z}=-\frac{Z}{2}\left(\alpha+\beta\right).
\end{equation}
The evaluation of the electronic repulsion integral, 
\begin{equation}
R_{\alpha,\beta}=\iint \frac{|\Psi_{\alpha,\beta}(\vec{r_1},\vec{r_2})|^2}{|\vec{r}_1-\vec{r}_2|} d^3 r_1 d^3 r_2
\end{equation} 
is a bit more challenging due to the presence of the boundary. However, it is still possible to give an analytical expression for $R$. For this purpose, we take the route via the multipole expansion, i.e. we write 
\begin{equation}\label{mpole}
\frac{1}{|\vec{r}_1-\vec{r}_2|}=\sum_{l=0}^{\infty}\frac{r_-^l}{r_+^{l+1}}P_l(\cos\theta_{12})
\end{equation}
where $r_+=\max\lbrace r_1, r_2\rbrace, r_{-}=\min\lbrace r_1, r_2\rbrace$ and $\theta_{12}$ is the angle between the vectors $\vec{r}_1,\vec{r}_2$. 
%Then,
%\begin{equation}
%R=\sum_{l=0}^{\infty}\iint_{r_1<r_2} 
%\end{equation}
This yields (see Appendix)
%\begin{equation}
%R_{\alpha,\beta}=8\pi^2 \sum_{l=0}^{\infty}\frac{c_l}{l+5} \left(J_l(2\alpha,2\beta)+J_l(2\beta,2\alpha)+2J_l(\alpha+\beta,\alpha+\beta)\right)
%\end{equation}
\begin{equation}\label{Ra}
R_{\alpha,\beta}=\frac{315\alpha^5\beta^5(\alpha+\beta)}{\pi^2\left((\alpha+\beta)^{10}+(2\sqrt{\alpha\beta})^{10}\right)} \sum_{l=0}^{\infty}\frac{c_lR_{l;\alpha,\beta}}{l+5}
\end{equation}
with 
\begin{equation}
c_l=\frac{\pi^3}{\left(4\left(\Gamma(\frac{4-l}{2})\Gamma(\frac{5+l}{2})\right)^2\right)}
\end{equation} 
where $\Gamma$ denotes the Euler gamma function, and 
\begin{equation}
R_{l;\alpha,\beta}=J_l(2\alpha,2\beta)+J_l(2\beta,2\alpha)+2J_l(\alpha+\beta,\alpha+\beta)
\end{equation}
with
\begin{equation}
J_l(x,y)={}_2F_1(9,1;l+6;\frac{y}{x+y})
\end{equation}
where ${}_2F_1$ is the hypergeometric function. The coefficients $c_l$ stem from the angular integral and decay very fast with increasing $l$. Note that in particular, all the even multipoles with $l>2$ vanish. 

\begin{figure}
\includegraphics{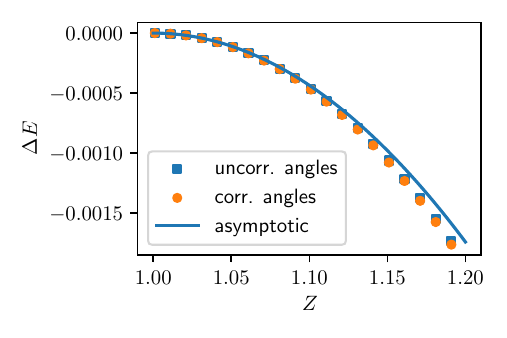}
\caption{Variational estimates of the binding energy $\Delta E$ as a function of $Z$, both from the pure Chandrasekhar Ansatz \eqref{ansatz} and its slight extension with explicit angular correlation, \eqref{angl}, together with the asymptotics close to the stability edge $Z=1$, Eq. \eqref{asm}. The existence of a bound state for $Z<1$ is excluded in the singlet sector by the lack of appropriate excited states for the hydrogen anion. }\label{ens}
\end{figure}

The minimization of the expected value $E_{\alpha,\beta}(Z)=T_{\alpha,\beta}+V_{\alpha,\beta,Z}+R_{\alpha,\beta}$ can now easily be done numerically. In Fig. \ref{ens} we plot the variational estimate of the binding energy so obtained, i.e. $\Delta E=E_{\alpha_Z,\beta_Z}(Z)+\frac{Z^2}{8}$, with $\alpha_Z,\beta_Z$ being the optimal values of the non--linear parameters given $Z$, and $\frac{Z^2}{8}$ is the ground state energy of $\mathbb{H}_1$, that is, for the ionized problem. By the HVZ theorem \cite{Teschl2014}, if $\Delta E<0$, we have binding in the system. As expected, we find $\Delta E<0$ for $Z>1$ and $\Delta E=0$ for $Z<1$, so that $Z=1$ is the critical value of $Z$ necessary for binding, in contrast to $Z=0.91...$ for the helium isoelectronic sequence in the full space problem. This suggests that one can effectively study the critical stability edge for anions using similar systems made of stacked semiconductor layers. We hence look more closely into this stability edge in our Ansatz in what follows.
\subsection{Stability edge and critical exponents}
One of the advantages of the simple Ansatz \eqref{ansatz} is that the stability edge $Z\rightarrow 1$ can be investigated by asymptotic analysis on account of the observation that the optimal value of either inverse effective Bohr radius, $\alpha$ or $\beta$, tends to zero as $Z$ approaches its critical value. Accordingly, we can expand $E_{\alpha,\beta}(Z)$ around zero to second order in the smaller variable, say, $\alpha$. The electron--hole potential energy has already the desired form; the kinetic energy reads, for $\alpha \ll 1$, 
\begin{equation}
T\approx \frac{1}{2}(\alpha^2 +\beta^2 ).
\end{equation}
To study the asymptotics of the electron repulsion, we use the formulae \cite{NIST:DLMF}
\begin{equation}\label{asymps}
\begin{cases}
 {}_2F_1(a,b;c;1)=\frac{\Gamma(c)\Gamma(c-a-b)}{\Gamma(c-a)\Gamma(c-b)}, & \mathrm{Re}(c-a-b)>0;\\
 \lim_{z\rightarrow 1^-}\frac{ {}_2F_1(a,b;c;1)}{-\ln(1-z)}=\frac{\Gamma(a+b)}{\Gamma(a)\Gamma(b)},  &c=a+b;\\
 \lim_{z\rightarrow 1^-}\frac{ {}_2F_1(a,b;c;1)}{(1-z)^{c-a-b}}=\frac{\Gamma(c)\Gamma(a+b-c)}{\Gamma(a)\Gamma(b)}, &  \mathrm{Re}(c-a-b)<0.
\end{cases}
\end{equation}
as well as ${}_2F_1(a,b;c,0)=1$; the latter implies that one can neglect $J_l(2\beta, 2\alpha)$ and $J_l(\alpha+\beta,\alpha+\beta)={}_2F_1(9,1,l+6,1/2)$ in comparison with $J_l(2\alpha, 2\beta)$ as $\alpha \ll 1$, while \eqref{asymps} shows 
\begin{equation}
J_l(2\alpha, 2\beta)\sim
\begin{cases}
\frac{\Gamma(l+6)\Gamma(l-4)}{\Gamma(l-3)\Gamma(l+5)}, & l>4 \\
-9\ln\frac{\alpha}{\beta}, & l=4\\
\left(\frac{\beta}{\alpha}\right)^{4-l}\frac{\Gamma(l+6)\Gamma(4-l)}{\Gamma(9)}, & l<4. 
\end{cases}
\end{equation}
Thus, since the prefactor multiplying the series in $R_{\alpha,\beta}$ behaves as $\sim \alpha^5\beta^{-5}$  we can limit ourselves to $l=0$ and $l=1$ terms in the sum if we are interested in the truncation to second order in $\alpha$. The corresponding hypergeometric functions are then given explicitly in terms of elementary functions and their relevant asymptotic expansions can be found easily: 
\begin{equation}
{}_2F_1(9,1;6,\frac{\beta}{\alpha+\beta})=\frac{\beta^4}{56\alpha^4}+\frac{9\beta^3}{56\alpha^3}+O(\frac{\beta^2}{\alpha^2})
\end{equation}
\begin{equation}
{}_2F_1(9,1;7,\frac{\beta}{\alpha+\beta})=\frac{\beta^3}{28\alpha^3}+\frac{9\beta^2}{28\alpha^2}+O(\frac{\beta}{\alpha})
\end{equation}
Combined with
\begin{equation}
\frac{\alpha^5\beta^5(\alpha+\beta)}{(\alpha+\beta)^{10}+(2\sqrt{\alpha\beta})^{10}}=\frac{\alpha^5}{\beta^4}-\frac{9\alpha^6}{\beta^4}+O(\frac{\alpha^7}{\beta^6})
\end{equation}
this gives
\begin{equation}
R_{\alpha,\beta}\approx
%\approx 4\pi^2 \frac{\alpha^5\beta^5}{\pi^2} \frac{8!}{2^9\beta^9}\left(\frac{c_0}{5}\left(\frac{\beta}{\alpha}\right)^{4}\frac{\Gamma(6)\Gamma(4)}{\Gamma(9)}+\frac{c_1}{6}\left(\frac{\beta}{\alpha}\right)^{3}\frac{\Gamma(7)\Gamma(5)}{\Gamma(9)}\right)=
\frac{1}{2}\alpha+\frac{15}{32}\frac{\alpha^2}{\beta}.
\end{equation}
Hence, for small $\alpha$, 
\begin{equation}\label{ap}
E(\alpha,\beta,Z)\approx\underbrace{\frac{1}{2}\beta^2-\frac{Z}{2}\beta}_{\mathrm{I}}+\underbrace{\frac{1}{2}\alpha^2-\frac{Z-1}{2}\alpha}_{\mathrm{II}}+\underbrace{\frac{15}{32}\frac{\alpha^2}{\beta}}_{\delta V^{\rm{eff}}_{\rm{corr}}}.
\end{equation}
The interpretation of the above is clear. The terms I and II describe a completely uncorrelated system composed of one electron and a hole of charge $Z$ and one electron and a hole of charge $Z-1$, which is in accordance with the simple picture of the farther electron (described by $\alpha$) being subject to a fully screened positive charge. It is the term $\delta V^{\rm{eff}}_{\rm{corr}}$ that correlates the electrons via the $\alpha/\beta$ dependence. This term originates from the $l=1$ component of the multipole expansion, and corresponds to a (repulsive) charge--dipole interaction. The emergence of the effective dipole here is due to the boundary, which induces a permanent dipole moment of the indirect exciton, whose orientation is such that it repels the other electron. In this way, the Chandrasekhar Ansatz nicely captures how the nature of the indirect exciton formation and electron correlation combine at the stability edge.

Further, it is easily seen that binding occurs only if $Z>1$, since otherwise $E(\alpha,\beta,Z)\geq \mathrm{I} \geq -Z^2/8$. The value of the critical charge $Z_c=1$ is determined here completely by the small $\alpha$ asymptotics of the $l=0$ term in the multipole expansion of the electron repulsion integral.

With \eqref{ap}, it is straightforward to find the critical behavior of the binding energy 
\begin{equation}\label{asm}
\Delta E=-\frac{(Z-Z_c)^2}{23}
\end{equation}
as well as the inverse effective Bohr radii
\begin{eqnarray}
\label{a}&\alpha-\alpha_c=\frac{4}{23}(Z-Z_c)\\ 
\label{b}&\beta-\beta_c=\frac{1}{2}(Z-Z_c)
\end{eqnarray}
where $\alpha_c=0,\beta_c=\frac{1}{2}, Z_c=1$. These quantities all exhibit power--law behavior under approaching the critical value of $Z$, in analogy to critical phenomena encountered in condensed matter systems, and thus the anion formation at stability edge can be interpreted as a second--order phase transition, as mentioned in \cite{Kais2000}.

The importance of electronic correlation in our setting is clearly manifest in the value of the coefficient in front of the term $(Z-1)^2$ in the binding energy. In fact, if the system was effectively uncorrelated at the stability edge, one would expect the behaviour $\Delta E= -\frac{(Z-1)^2}{8}$ describing the binding energy of the farther electron coupled to the screened nucleus of effective charge. This is what one observes in the behavior of the energy of the excited bound state of the helium isoelectronic sequence \cite{Pachucki2012}. The reduction of this coefficient from $1/8$ to $1/23$ is a consequence of the correlation of the electrons beyond the simple screening effect, which, as we see, is quite significant.

\section{Discussion}
\label{sec:discuss}
\subsection{Radial vs. angular correlation}
\begin{figure}
\includegraphics{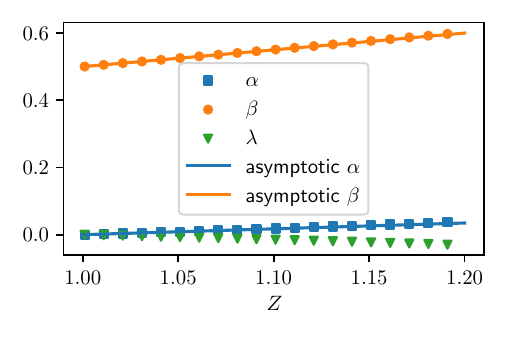}
\caption{Optimal values of the variational parameters $\alpha,\beta,\lambda$ in the extended Chandrasekhar Ansatz \eqref{angl}. The asymptotic behavior is described by Eqs. \eqref{a} and \eqref{b}. Note that while $\alpha$ and $\beta$ are measured in units of the inverse effective Bohr radius, $\lambda$ is dimensionless.} \label{abc}
\end{figure}

We need to comment on the limitations of our approach. Despite its simplicity, the Chandrasekhar wave function is good enough to capture the fact that $Z=1$ is critical. One might ask how well does it reproduce the critical behavior.  In order to estimate the effects of angular correlation between the electrons at the stability edge, we extended the trial state slightly by taking the trial function 
\begin{equation}\label{angl}
\Psi(r_1,r_2;\alpha,\beta;\lambda)=N_{\lambda} \left(1+\lambda \cos\theta_{12}\right)\Psi_{\alpha,\beta}(\vec{r_1},\vec{r_2})
\end{equation}
where $\theta_{12}$ is the angle between $\vec{r}_1$ and $\vec{r}_2$, $\lambda$ is a linear parameter quantifying the angular correlation, with $N_{\lambda}$ being the normalization factor. In Figs. \ref{ens} and \ref{abc}  we plot the resulting energies, in comparison with the simple Chandrasekhar Ansatz, and the optimal values of $\lambda$, respectively. We observe that angular correlation at this simple level becomes increasingly less important as one approaches the stability edge, with $\lambda$ negative -- in accordance with the expectation that the angular motion of the electrons is anticorrelated -- and tending to zero as $Z$ approaches its critical value. In particular, the asymptotic behavior of the binding energy at the stability edge remains the same as obtained from the Chandrasekhar function. Of course, a more complete treatment of the problem beyond the very simple trial function we use is possible with extensive basis sets, e.g. of gaussian or Hylleraas type. We do not attempt to provide such exact calculations at this point. It is to be expected, nevertheless, that they can be effectively checked against our results at the stability edge, in particular the power--law expressions \eqref{asm}, \eqref{a} and \eqref{b}.

\subsection{Continuous tuning of $Z$ in experiment}
Beyond the obvious limitations of the variational wave function, when describing the arrangement in question it is crucial to take into account the finite layer thickness as well as the finite mass of the hole which would be present in any realistic setting. While the second aspect appears to disfavor the binding (see the phase diagram in~\cite{Kais2000}), the finite size effects, combined with dielectric corrections, should in turn act towards stabilizing the trion state, as, for instance, the $2d$ trion is bound \cite{Witham2018}. The impact of dielectric effects due to the bound charges forming at the boundaries is thus important and we are going to discuss it very briefly. Assume that the dielectric constant of the electron--rich layer, $\varepsilon_1$, is larger than the dielectric constant $\varepsilon_2$ of the material present below the boundary containing the hole: $\varepsilon_1>\varepsilon_2$. Then the potential generated by the hole of unit charge is everywhere equal to \cite{Batygin1978}
\begin{equation}
V_h(r)=-\frac{2}{\varepsilon_1+\varepsilon_2}\frac{1}{r}\equiv -\frac{Z_{\mathrm{eff}}}{\varepsilon_1r}
\end{equation}
where $Z_{\mathrm{eff}}>1$ since $\varepsilon_1>\varepsilon_2$.
Now if we disregard the image charges generated by the electrons, their mutual interaction equals $1/(\varepsilon_1 |\vec{r}_1-\vec{r}_2|)$ and the hamiltonian of the system maps onto $\mathbb{H}_2$ with an effective charge $Z$ which can be varied \emph{continuously} by tuning $\varepsilon_1/\varepsilon_2$, in particular $Z-Z_c\approx \frac{\varepsilon_1-\varepsilon_2}{\varepsilon_1}$ and thus the stability edge corresponds to identical materials placed above and below the surface hosting the hole. In this scenario, not only the stability edge appears accessible, but also can be approached continuously by proper design of the material. The mapping onto $\mathbb{H}_2$ is of course not exact because of the bound charges due to the electrons, but the onset of a continuous effective $Z$ of the hole appears plausible in the study of indirect excitonic complexes so formed. %Under these circumstances, one could study not only the approach to the stability edge but also the ion--to--atom transition, which manifests itself in the %This provides another argument in favor of using indirect excitons as a platform to simulate critically bound anions. 

\subsection{Anion--to--atom transition}
\begin{figure}
\includegraphics{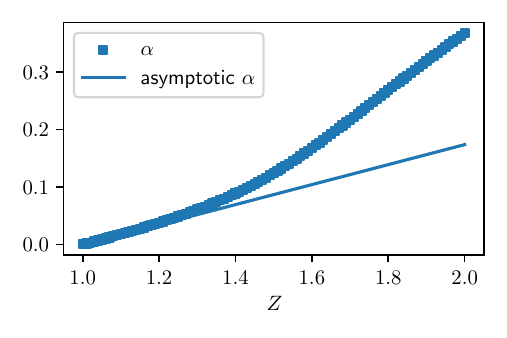}
\caption{Behavior of the optimal value of the variational parameter $\alpha$ determining the effective inverse Bohr radius of the \emph{further} electron, as a function of $Z$. A crossover from the asymptotic regime of a weakly bound ion described by Eq. \eqref{a} to the quasi--neutral regime around $Z=2$ is observed.}\label{crossA}
\end{figure}

\begin{figure}
\includegraphics{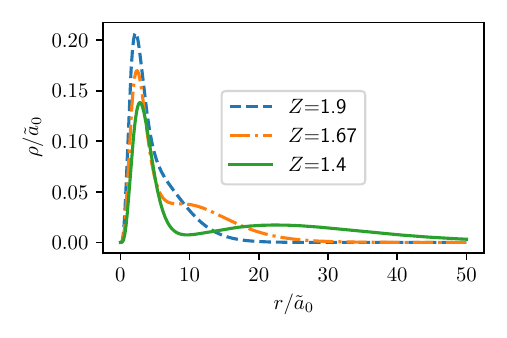}
\caption{Chandrasekhar radial probability density of finding an electron at distance $r$ from the hole, integrated over angles. Close to neutrality, it possesses a single maximum: for lower $Z$, a second maximum appears, whose position wanders off to infinity as $Z\rightarrow 1$. The number of maxima can be employed to characterize the system as atom--like and or anion--like. $\tilde{a}_0$ is the Bohr radius corresponding to electrons with the appropriate effective mass and the dielectric constant of the material, equal to unity in the pseudo--atomic units $\hbar=1, m_e=1$ used in the text. }\label{radP}
\end{figure}

With the plausible option of tuning $Z$ continuously by material design, another interesting possibility arises in our context. In \cite{Kais2000}, the phase diagram of three--body Coulomb systems under variations of masses and charges was projected, with one of its conclusions being the classification of these systems into \emph{atom--like}, e.g. He, and \emph{molecule--like}, e.g. $\mathrm{H}_2^+$. In our case, if we move away from the vicinity of the critical charge $Z=1$ and look at the entire range of $Z$ corresponding to excess charge in our simple Chandrasekhar--type Ansatz, we observe that the optimal values of the smaller non--linear parameter $\alpha$ display a crossover from the regime where its dependence of $Z$ fits the asymptotic linear behavior close to criticality, Eq. \eqref{a}, and another linear behavior around $Z=2$ (Fig. \ref{crossA}). This behavior can be attributed to the ion--to--atom transition similar to the one found in the study \cite{Kais2000}; in fact, when the radial position probability density is plotted, Fig. \ref{radP}, one observes that it either possesses two distinct maxima, corresponding to an anion--like structure with one of the electrons well--separated from the core, or a single maximum, as can be expected from the charge density of the helium--like atom (here, in a $P$ state), with the second minimum disappearing for $Z\approx 1.7$.  If $Z$ can be indeed varied continuously by using indirect trions formed in appropriately devised materials, additional insight into this particular transition could arise, beyond the study of the critical binding mechanisms. 

\section{Summary and outlook}
We have discussed the problem of the existence of a three--body bound state composed of two light negatively charged particles and a single immobile positive charge in the geometric setting where the electrons are confined to move in common half--space bounded by the surface hosting the positive charge. Using a Chandrasekhar--type trial state, we found that the critical charge needed to bind the two negative particles equals exactly unity, in contrast to the case of the helium isoelectronic sequence where the critical $Z$ equals approximately $0.911$. We also discussed the critical behavior of the system close to the stability edge and identified the critical exponents and amplitudes governing the binding energy and effective Bohr radii. Asymptotically, the electronic correlation effects manifest themselves as an effective  charge--dipole repulsion. Since the setting considered can be thought of an idealized model of an interlayer trion where the electron--rich layer is of a much larger thickness than the adjacent layer hosting heavy holes, we argue that indirect excitonic complexes of this kind can be employed as an experimental platform to study anions at the critical charge necessary for binding. Our study connects to basic questions in anion chemistry related to electronic correlation, such as the well--known maximal ionization problem. It has a broad potential for extensions, both in the direction of performing precise energy calculations beyond the Chandrasekhar Ansatz and, most importantly, by taking the finiteness of the layers and the mobility of the hole into account, together with a complete treatment of the dielectric effects. All these aspects are definitely within reach with the use of current methods in computational many--body physics and chemistry. 

\emph{Acknowledgements.} We thank Grzegorz Łach for inspiring discussions. This work was supported by the National Science Centre of Poland Grant 2020/37/B/ST2/00486.

%\bibliography{refs2}

\appendix*
\begin{widetext}
\section{Evaluation of the repulsion integral}
The repulsion integrals involving Slater--type orbitals, as in our case, are typically easily computed from \emph{master integral formulae}, cf. \cite{Pachucki2009}. In our case, the presence of the boundary makes these formulae inapplicable; nevertheless, the repulsion integral can still be given in a relatively simple closed form. 
As announced in the main text, we start from the multipole expansion \eqref{mpole}, which casts the repulsion integral into the form
\begin{equation}
R_{\alpha,\beta}=2\sum_{l=0}^{\infty}\iint_{r_1<r_2} d^3 \vec{r}_1 d^3 \vec{r}_2 |\Psi_{\alpha,\beta}(\vec{r_1},\vec{r_2})|^2 \frac{r_1^lP_l(\cos{\theta}_{12})}{r_2^{l+1}}
\end{equation}
where we used the symmetry of the wave function under the exchange of coordinates. Employing this again, together with the separation of $\Psi$ into the angular and radial part, we obtain 
%\begin{widetext}
\begin{equation}
R_{\alpha,\beta}=\frac{4}{\pi^2}\frac{\alpha^{5}\beta^{5}}{1+\left(\frac{2\sqrt{\alpha\beta}}{\alpha+\beta}\right)^{10}}\sum_{l=0}^{\infty}c_l \left(\tilde{J}_l(2\alpha,2\beta)+\tilde{J}_l(2\beta,2\alpha)+2\tilde{J}_l(\alpha+\beta,\alpha+\beta)\right)
\end{equation}
%\end{widetext}
where 
\begin{equation}
\tilde{J}_l(x,y)=\int_0^{\infty}dt ~t^{l-3}e^{-xt} \int_0^t ds ~s^{4+l} e^{-ys}
\end{equation}
and 
%\begin{widetext}
\begin{equation}
c_l=\int_0^{2\pi} d\varphi_1 \int_0^{2\pi} d\varphi_2 \int_0^{\pi/2} d\theta_1 \int_0^{\pi/2} d\theta_2 \cos^2\theta_1 \cos^2\theta_2 \sin \theta_1 \sin\theta_2 P_l(\cos\theta_{12})
\end{equation}
%\end{widetext}
where we used the standard spherical coordinates for the electrons and taken into account the fact that $\Psi=0$ for $\theta_1,\theta_2\geq\pi/2$. To evaluate the angular integral, we use 
\begin{equation}
\begin{split}
&P_l(\cos\theta_{12})=P_l(\cos\theta_1)P_l(\cos\theta_2)+\\
&+2\sum_{m=1}^l\frac{(l-m)!}{(l+m)!}P^m_l(\cos\theta_1)P^m_l(\cos\theta_2)\cos(m(\varphi_1-\varphi_2);
\end{split}
\end{equation}
the entire sum over $m$ vanishes due to the $\varphi_1,\varphi_2$ integration and we are left with 
\begin{equation}
c_l=4\pi^2\left[\int_0^1 x^2 P_l(x) dx\right]^2.
\end{equation}
The integral involving the Legendre polynomial can be evaluated using the recurrence formula
\begin{equation}
(l+1)P_{l+1}(x)-(2l+1) x P_l(x)+l P_{l-1}(x)=0
\end{equation}
twice, together with
\begin{equation}
\int_0^1P_l(x) dx = 
\begin{cases}
\frac{1}{2}, & l=0 \\
\left(-\frac{1}{2}\right)^{l(l-1)/2}\frac{(l-2)!!}{2\Gamma(\frac{l+1}{2})}, & l  \text{   odd}\\
0, & l  \text{    even}.
\end{cases}
\end{equation}
which follows from parity, the orthogonality relation for the Legendre polynomials as well as the the Rodrigues formula. It yields 
\begin{equation}
c_l=\frac{\pi^3}{4\left(\Gamma(\frac{4-l}{2})\Gamma(\frac{5+l}{2})\right)^2}
\end{equation}
as given in the main text. On the other hand, the radial integral can be written 
\begin{equation}
\tilde{J}_l(x,y)=\frac{1}{y^9}\int_0^{\infty} dt ~ t^{3-l}\gamma(5+l,t) e^{-\frac{x}{y}t}
\end{equation}
with $\gamma(q,t)$ denoting the lower incomplete Gamma function. Expressing the latter via the power series 
\begin{equation}
\gamma(q,t)=t^q \Gamma(q)e^{-t}\sum_{k=0}^{\infty}\frac{t^k}{\Gamma(q+k+1)}
\end{equation}
and integrating, we obtain 
\begin{equation}
\begin{split}
&\tilde{J}_l(x,y)=\frac{\Gamma(5+l)}{(x+y)^9}\sum_{k=0}^{\infty}\frac{\Gamma(9+k)}{\Gamma(6+l+k)}\left(\frac{y}{y+x}\right)^k\\
&=\frac{\Gamma(9)}{(x+y)^9(l+5)}\left[{}_2F_1(9,1;l+6;\frac{y}{y+x})\right]
\end{split}
\end{equation}
where we used the definition of the hypergeometric function
\begin{equation}
{}_2F_1(a,b;c;z)=\frac{\Gamma(c)}{\Gamma(a)\Gamma(b)}\sum_{k=0}^{\infty} \frac{\Gamma(a+k)\Gamma(b+k)}{\Gamma(k+1)\Gamma(c+k)}z^k.
\end{equation}
Combining these results, we arrive at the series representation \eqref{Ra} given in the main text. 
\end{widetext}
\bibliography{refs2}
\end{document}